\documentclass[twoside,slac_one]{revtex4}
\usepackage{graphicx}
\usepackage{fancyhdr}
\usepackage{amsmath} 
\usepackage{bm}
\usepackage{amsxtra}
\usepackage{amssymb}
\usepackage{amsthm}
\usepackage{latexsym}
\usepackage{lscape}

\def\cp{$CP$\/}

\def\mevm{~MeV/$c^2$\/}
\def\mevp{~MeV/$c$\/}
\def\meve{~MeV}
\def\gevm{~GeV/$c^2$\/}

\def\geve{~GeV}

\def\ra{\!\rightarrow\!}

\def\bbar{\overline{B}{}^{\,0}}

\def\bsdsds{$B^0_s\ra D^{(*)+}_s D^{(*)-}_s$}
\def\bsdspi{$B^0_s\ra D^{(*)-}_s \pi^+$}
\def\bdsd{$B^0\ra D^{(*)+}_s D^-$}

\def\bs{B^{}_s}
\def\bsst{B^{*}_s}
\def\bsbar{\overline{B}{}^{}_s}
\def\bsbarst{\overline{B}{}^{\,*}_s}
\def\fl{f^{}_L}

\def\mbc{M^{}_{\rm bc}}
\def\de{\Delta E}
\def\qq{$q\bar{q}$}

\def\dgs{\Delta\Gamma^{}_s}
\def\dgcp{\Delta\Gamma^{CP}_s}
\def\gs{\Gamma^{}_s}
\def\kstz{\overline{K}{}^{\,*0}}
\def\kstp{K^{*+}}
\def\qqbar{$q\bar{q}$}

\begin{document}

\title{Updated Measurement of ${\cal B}(B_s \to D_s^{(*)+}D_s^{(*)-})$  and Determination of $\Delta \Gamma_{s}$ }

%

\author{S. Esen}
\affiliation{Department of Physics, University of Cincinnati, Cincinnati, OH, USA}

\begin{abstract}
Using fully reconstructed $B_{s}$ mesons, we measure exclusive branching fractions for the decays $B_s \to D_s^{(*)+}D_s^{(*)-}$. 
The results are 
${\cal B}(B^0_s\ra D^+_s D^-_s)=(0.58\,^{+0.11}_{-0.09}\,\pm 0.13)\%$,
${\cal B}(B^0_s\ra D^{*\pm}_s D^{\mp}_s)=(1.8\, \pm 0.2\,\pm 0.4)\%$, 
and
${\cal B}(B^0_s\ra D^{*+}_s D^{*-}_s)=(2.0\,\pm 0.3\,\pm 0.5)\%$;
the sum is 
${\cal B}(B^0_s\ra D^{(*)+}_s D^{(*)-}_s)=(4.3\,\pm 0.4\,\pm 1.0)\%$.
Assuming these decay modes saturate decays to CP-even final states, the branching fraction determines the relative width difference between the $B_s$ $CP$-odd and $CP$-even eigenstates. 
Taking \cp\ violation to be negligibly small, we obtain $\dgs/\gs = 0.090\,\pm 0.009\,{\rm(stat.)}\,\pm 0.022 \,{\rm (syst.)}$, where $\gs$ is the mean decay width. The results are based on a data sample collected with the Belle detector at the KEKB $e^+ e^-$ collider running at the $\Upsilon(5S)$ resonance with an integrated luminosity of 121.4 fb$^{-1}$. 
\end{abstract}

\maketitle

\thispagestyle{fancy}


\section{Introduction}

An $e^+e^-$ collider running at a center-of-mass (CM) energy corresponding to the $\Upsilon(5S)$ resonance can produce significant amounts of $B^{(*)}_s\overline{B}{}^{(*)}_s$ pairs~\cite{bs_cleo,drutskoy07}. The Belle detector~\cite{belle_detector} at the KEKB asymmetric-energy $e^+ e^-$ collider~\cite{kekb} has collected a data sample corresponding to an integrated luminosity of 121.4 fb$^{-1}$  at the $\Upsilon(5S)$ resonance ($\sqrt{s}=10.87$\geve). This sample has allowed us to make the most precise measurement of  the branching fractions of  \bsdsds decays~\cite{charge-conjugates}. These Cabibbo-favored final states are expected to be predominantly \cp-even~\cite{Aleksan} and dominate the difference in decay widths $\dgcp$ between the two $\bs$-$\bsbar$ \cp\ eigenstates~\cite{Aleksan}. We report preliminary results of the updated branching fraction measurements, which replace our previous measurement based on 23.6 fb$^{-1}$ of data~\cite{esen10}.  

\section{Event Selection}

The Belle detector consists of a silicon vertex detector (SVD), a 50-layer central drift chamber (CDC), an array of aerogel threshold Cherenkov counters (ACC), time-of-flight scintillation counters (TOF), and an electromagnetic calorimeter (ECL) comprising of CsI(Tl) crystals located inside a superconducting solenoid coil that provides a 1.5 T magnetic field. 
For charged hadron identification, a likelihood ratio is formed based on a dE/dx measurement in the CDC and the response of the ACC and TOF. 
Only good quality charged tracks originating from near the $e^+e^-$ interaction point are accepted. For charged kaon tracks, this likelihood ratio is required to be $>\!0.60$; the tracks not satisfying this requirement are identified as pions.   The kaon likelihood requirement is $\sim\!90$\% efficient and has a $\pi^\pm$ misidentification rate of $\sim\!10$\%.
With the exception of the tracks originating from $K^0_S$ decays, low-momentum charged tracks with $P<\!100$\mevp\ are rejected. Neutral $K^0_S$ candidates are reconstructed from $\pi^+\pi^-$ pairs having an invariant mass within 10\mevm\ of the nominal $K^0_S$ mass~\cite{pdg} and satisfying momentum-dependent requirements based on the decay vertex position~\cite{goodKS}. 

Neutral $\pi^0$ candidates are reconstructed from $\gamma\gamma$ pairs having an invariant mass within 15\mevm\ of the $\pi^0$ mass with each photon having a laboratory energy greater than 100\meve. 
Neutral and charged $K^{*}$ candidates are reconstructed from a $K$ and $\pi^+$ having an invariant mass within 50\mevm\ of $M^{}_{K^{*}}$. 
Neutral $\phi$ candidates are reconstructed from $K^+K^-$  pairs having an invariant mass within 12\mevm\  of $M^{}_{\phi}$.
 Charged $\rho^+$ candidates are reconstructed from $\pi^+\pi^0$ pairs having an invariant mass within 100\mevm\  of $M^{}_{\rho^+}$.

 We reconstruct $D^+_s$ candidates using six final states:  $\phi\pi^+$,
$K^0_S\,K^+$,
$\kstz K^+$,
$\phi\rho^+$,
$K^0_S\,\kstp$, and 
$\kstz \kstp$.
The invariant mass windows used are 10\mevm\ ($\sim\!3\sigma$) for the three final states containing $K^*$ candidates,  20\mevm\ ($2.8\sigma$) for $\phi\rho^+$, and 15\mevm\ ($\sim\!4\sigma$) for the remaining two modes. For the three vector-pseudoscalar final states, we require $|\cos\theta^{}_{\rm hel}|>0.20$, where the helicity angle $\theta^{}_{\rm hel}$ is the angle between the momentum of the charged daughter of the vector particle and the direction opposite the $D_s$ momentum, in the rest frame of the vector particle. 

We combine $D^+_s$ candidates with photon candidates to reconstruct $D^{*+}_s\ra D^+_s\gamma$ decays, and we require that the mass difference $M^{}_{\tilde{D}_s^+\gamma} - M^{}_{\tilde{D}_s^+}$ be within 12.0\mevm\ of the nominal value (143.8\mevm), where $\tilde{D}_s^+$ denotes the reconstructed $D^+_s$ candidate. This requirement (and that for the  $D^+_s$ mass) is determined by optimizing a figure-of-merit $S/\sqrt{S+B}$, where $S$ is the expected signal based on Monte Carlo (MC) simulation and $B$ is the expected background as estimated from $D^+_s$ sideband data. We require that the photon energy in the CM system be greater than 50\meve, and that the energy deposited in the central $3\times 3$ array of cells of the ECL cluster contain at least 85\% of the energy deposited in the central $5\times 5$ array of cells.

We select $B^0_s\ra D^{*+}_s D^{*-}_s$, $D^{*\pm}_s D^{\mp}_s$,  and $D^{+}_s D^{-}_s$ decays using two quantities evaluated in the $e^+e^-$ CM frame: the beam-energy-constrained mass $\mbc=\sqrt{E^2_{\rm beam} - p^2_B}$, and the energy difference $\de= E^{}_B-E^{}_{\rm beam}$, where $p^{}_B$ and $E^{}_B$ are the reconstructed momentum and energy of the $B^0_s$ candidate, and $E_{\rm beam}$ is the beam energy. 
We determine our signal yields by fitting events in the region $5.25\mbox{\gevm}<\mbc <5.45$\gevm\ and $-0.15\mbox{\geve}<\de <0.10$\geve.
Within this region, the modes $\Upsilon(5S)\ra\bs\bsbar$, $\bs\bsbarst$ and $\bsst\bsbarst$ are well-separated as the $\gamma$ from $\bsst\ra\bs\gamma$ decay is not reconstructed. We expect only small amounts of signal in $\bs\bsbar$ and $\bs\bsbarst$ and thus do not use these modes for the branching fraction measurement. In order to simplify the fit, we fix their relative ratios, which are determined from fully reconstructed $B_s^0 \to D_s^- \pi^+$ decays as described in  Ref.~\cite{remi}. 

When multiple \bsdsds\ candidates are reconstructed in an event, we select the candidate that minimizes the quantity 
\begin{eqnarray}
\chi^2 & = & \frac{1}{(2+N)}\,\biggl\{
\sum_{\#D^{}_s} \left[ \frac{(\tilde{M}^{}_{D_s} - M^{}_{D_s})}{\sigma^{}_M}\right]^2 
+ \sum_{\#D_s^{*}} \left[ \frac{(\widetilde{\Delta M} - 
\Delta M)}{\sigma^{}_{\Delta M}}\right]^2\biggr\}\,,
\end{eqnarray}
where $\Delta M=M^{}_{D^*_s}-M^{}_{D^{}_s}$, the quantities $\tilde{M}^{}_{D^{}_s}$ and $\widetilde{\Delta M}$ are reconstructed, and the summations run over the two $D^+_s$ daughters and the possible $D^{*+}_s$ daughters ($N\!=\!0,1,2$) of a $B^0_s$ candidate. 
 The mean mass $M^{}_{D^{}_s}$ and widths $\sigma^{}_M$ and $\sigma^{}_{\Delta M}$ are obtained from MC simulation and calibrated for any data-MC difference using a  $B^0\ra D^{(*)+}_sD^-$ sample in 563 $fb^{-1}$ of data at the $\Upsilon(4S)$ energy. Approximately half of the events have multiple candidates according to MC simulation, and this criterion selects the correct $B^0_s$ candidate 83\%, 73\%, and 69\% of the time for $D^+_s D^-_s$, $D^{*\pm}_s D^{\mp}_s$, and $D^{*+}_s D^{*-}_s$ final states, respectively.

The background from $e^+e^-\ra q\bar{q}~(q=u,d,s,c)$ continuum events is rejected using a  Fisher discriminant based on a set of modified Fox-Wolfram  moments~\cite{KSFW}. This discriminant distinguishes jet-like \qq\ events from more spherical $B^{}_{(s)}\overline{B}{}^{}_{(s)}$ events, and is used to calculate a likelihood $\mathcal{L}_{s}$ ($\mathcal{L}_{q\overline q}$) for an event assuming the event is signal ($q\overline{q}$ background). We require the ratio $\mathcal{R}=\mathcal{L}_{s}/(\mathcal{L}_{s}+\mathcal{L}_{q\overline q})$ to be $>\!0.20$. This selection is 93\% efficient for signal and removes $>62$\% of \qqbar\ background.
The majority of the background consists of 
$\Upsilon(5S)\ra B^{(*)}_s\overline{B}{}^{(*)}_s\ra D^+_s X$; $\Upsilon(5S)\ra BBX$ (where $b\bar{b}$ hadronizes into  $B^0,\,\bbar$, or $B^\pm$); and
$B^{}_s\ra D^\pm_{sJ}(2317)D^{(*)}_s$,
$B^{}_s\ra D^\pm_{sJ}(2460)D^{(*)}_s$, and
$B^{}_s\ra D^\pm_{s}D^\mp_s\pi^0$ decays. The last three processes peak at negative values of $\de$, and their yields are expected to be very small assuming their branching fractions are similar to analogous $B^{}_d\ra D^\pm_{sJ}D^{(*)}$ decays.  

Signal yields are measured using a two-dimensional extended unbinned maximum-likelihood fit to the $\mbc$-$\de$ distributions. For each signal decay, we include probability density functions (PDFs) for signal and background. 
We use a single PDF for background which consists of $q\bar{q}$, $B^{(*)}_s\overline{B}{}^{(*)}_s\ra D^+_s X$, and $\Upsilon(5S)\ra BBX$ events. The background PDF is constructed using an ARGUS function~\cite{ARGUS} for $\mbc$ and a second-order Chebyshev polynomial for $\de$.  The two parameters of the Chebyshev function are taken from the data in which one of the $D_s$ ``candidates" is required to be within the mass sideband.
The signal PDFs have three components: correctly reconstructed (CR) decays; ``wrong combination'' (WC) decays in which a non-signal track or photon is included in place of a true daughter track or photon; and ``cross-feed'' (CF) decays in which a $D^{*\pm}_s D^{\mp}_s$ or $D^{*+}_s D^{*-}_s$ is reconstructed as a $D^+_s D^-_s$ or $D^{*\pm}_s D^{\mp}_s$, respectively, or else a $D^+_s D^-_s$ or $D^{*\pm}_s D^{\mp}_s$ is reconstructed as a $D^{*\pm}_s D^{\mp}_s$ or $D^{*+}_s D^{*-}_s$. 
For these CF candidates $\de$ is shifted by 100-150\meve, but $\mbc$ remains almost unchanged. 
When the $B^0_s$ is not fully reconstructed, e.g. due to losing the $\gamma$ from $D^{*+}_s\ra D^+_s\gamma$ (CF-down), a negative shift in $\de$ is observed.   
Conversely, in the case where the signal decay has gained a photon (CF-up), $\de$ is typically shifted higher. 
The PDF for CR events is modeled with a  Gaussian for $\mbc$ and a double Gaussian with common mean for $\de$. CF and WC events have more complicated distributions. All signal shape parameters are taken from MC and calibrated using \bsdspi\ and \bdsd\ decays. The fractions of WC and CF-down events are taken from MC simulation. The fractions of CF-up events are difficult to calculate accurately from MC simulation as not all $B_s^0$ partial widths are measured; thus they are allowed to vary in the fit. As the CF-down fractions are fixed, the three distributions ($D^+_s D^-_s,\ D^{*\pm}_s D^{\mp}_s$, and $D^{*+}_s D^{*-}_s$) are fitted simultaneously~\cite{fit_errors}. The CF fractions are listed in Table~\ref{tab:fractions}.

\begin{table*}[ptbh]
\caption{ Fractional distribution of the signal reconstruction types from MC simulation of $B_s^0$ decay modes.  }
\begin{center}
\renewcommand{\arraystretch}{1.4}
\begin{tabular*}{.9\textwidth}{@{\extracolsep{\fill}} l | c c  c c @{\extracolsep{\fill}} }
\hline
\hline 
 
 $B_s^0$ Mode   &  RC & WC & CF I & CF II \\[0.5ex]
\hline 
  $D^+_s D^-_s$			 & 76.1 	& 6.0 fixed	 	& 17.1  ($\ra D^{*\pm}_s D^{\mp}_s$) & 0.8 ( $\ra D^{*+}_s D^{*-}_s$)\\[0.5ex]
$D^{*\pm}_s D^{\mp}_s$	& 44.4	& 38.5 fixed		& 8.2 ($\ra D^+_s D^-_s$)fixed	  	& 8.9 ( $\ra D^{*+}_s D^{*-}_s$)  \\[0.5ex]
 $D^{*+}_s D^{*-}_s$		&  31.8  	& 37.6 fixed		& 2.0 ($\ra D^+_s D^-_s$) fixed  	 & 28.6 ($\ra D^{*\pm}_s D^{\mp}_s$) fixed	\\[0.5ex]

\hline
\hline
\end{tabular*}
\end{center}
\label{tab:fractions}
\end{table*}

\subsection{Results}

\begin{figure}[ptbh]
\centering
\includegraphics[width=55mm]{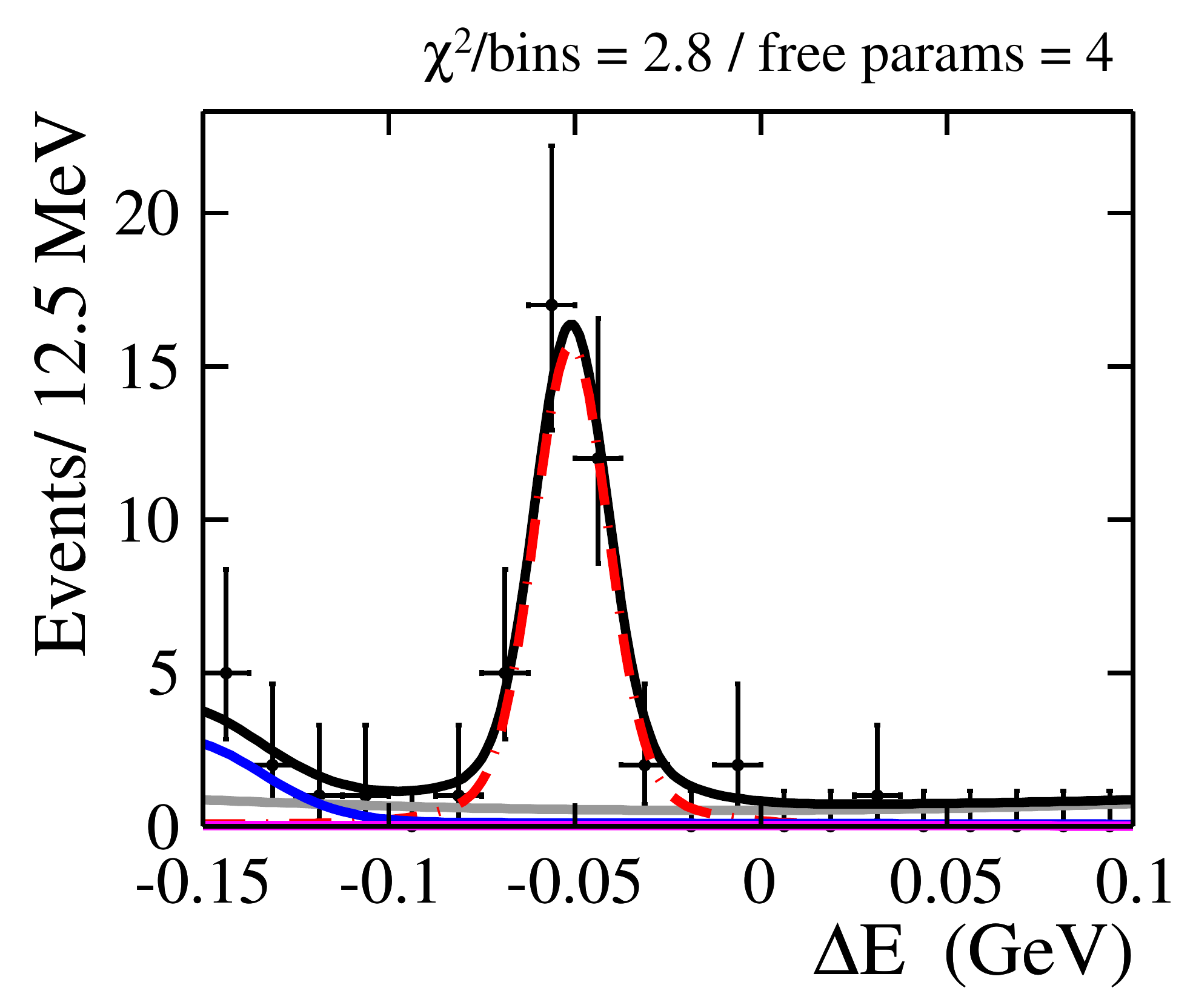}
\includegraphics[width=55mm]{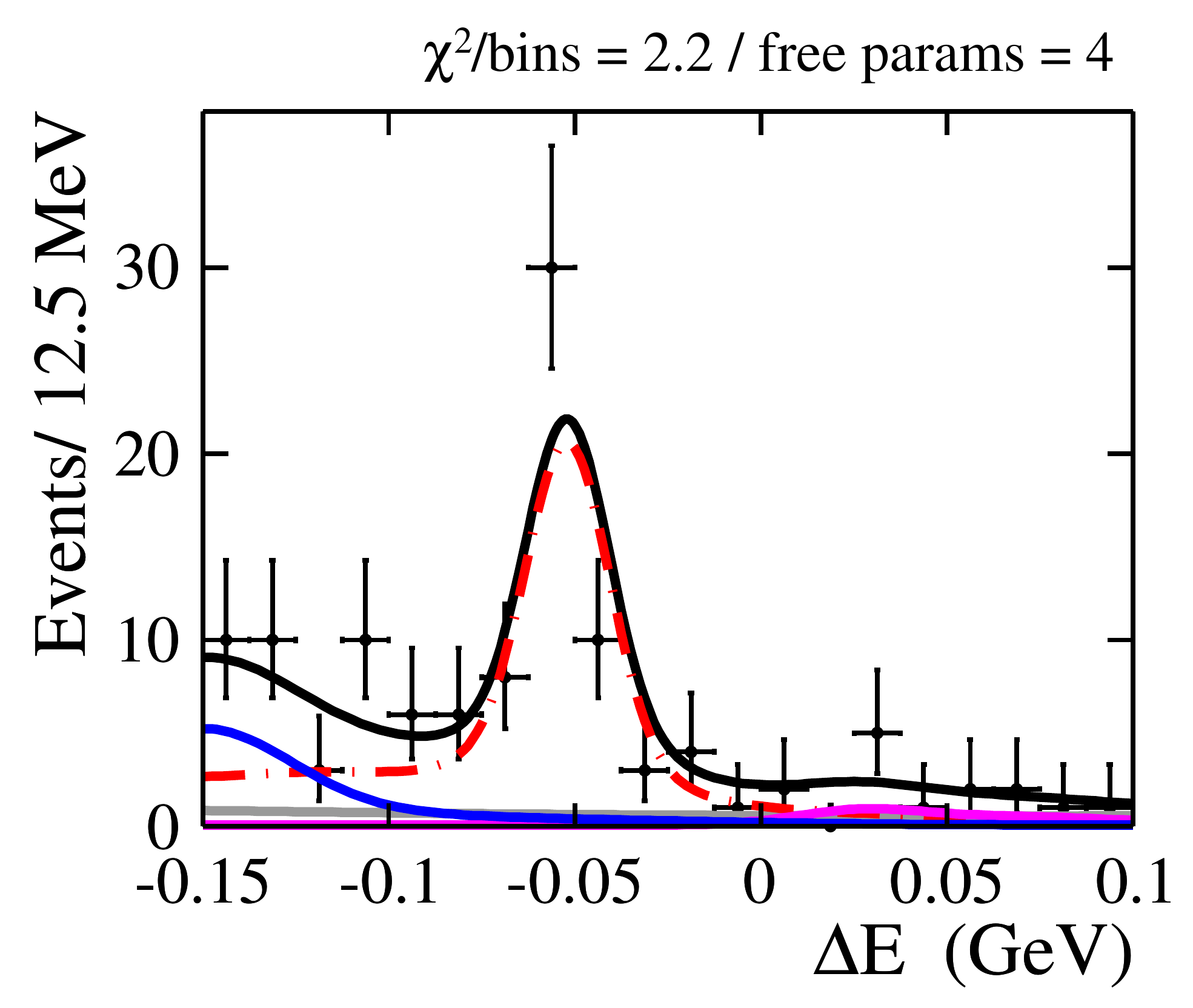}
\includegraphics[width=55mm]{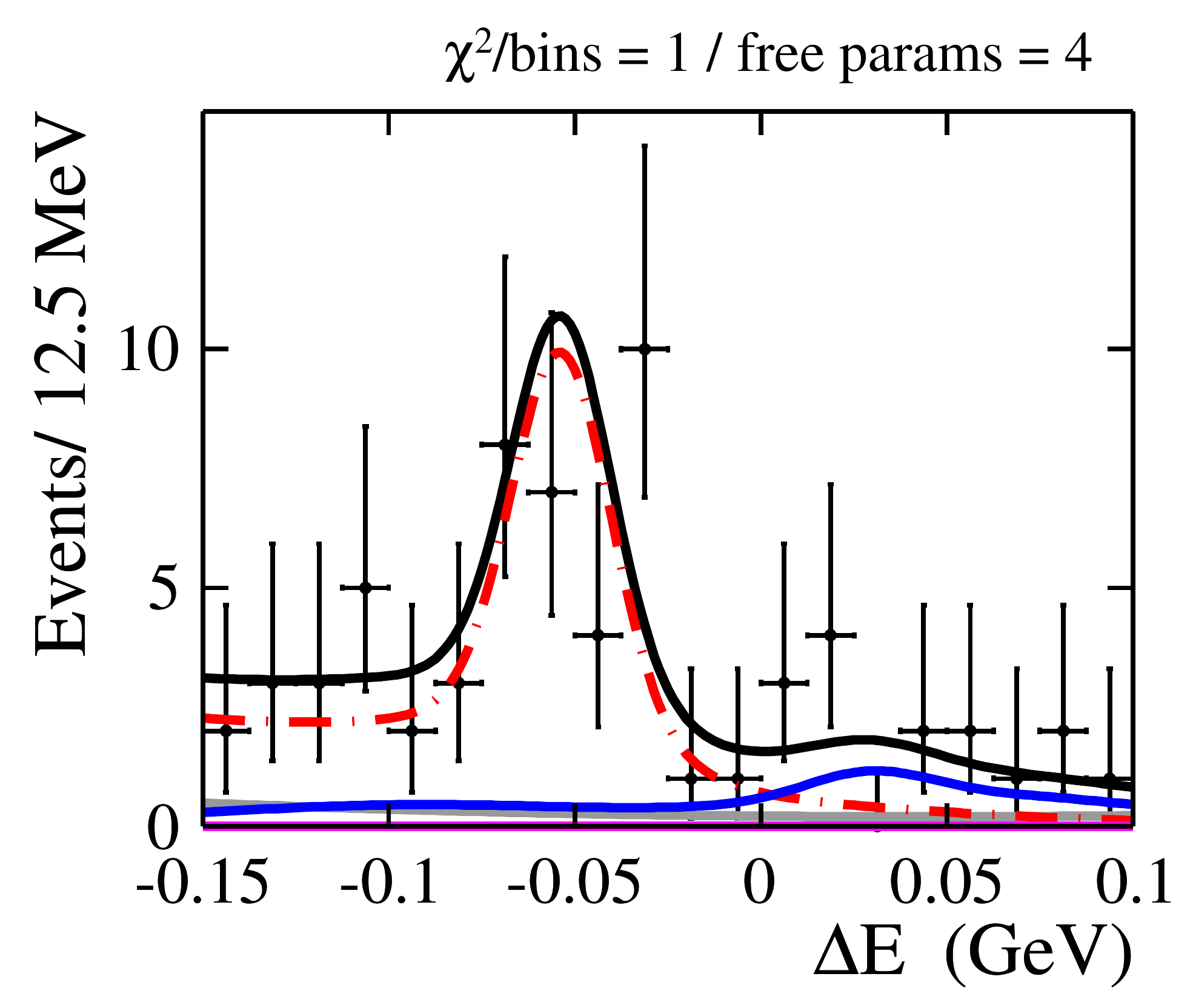}

\includegraphics[width=55mm]{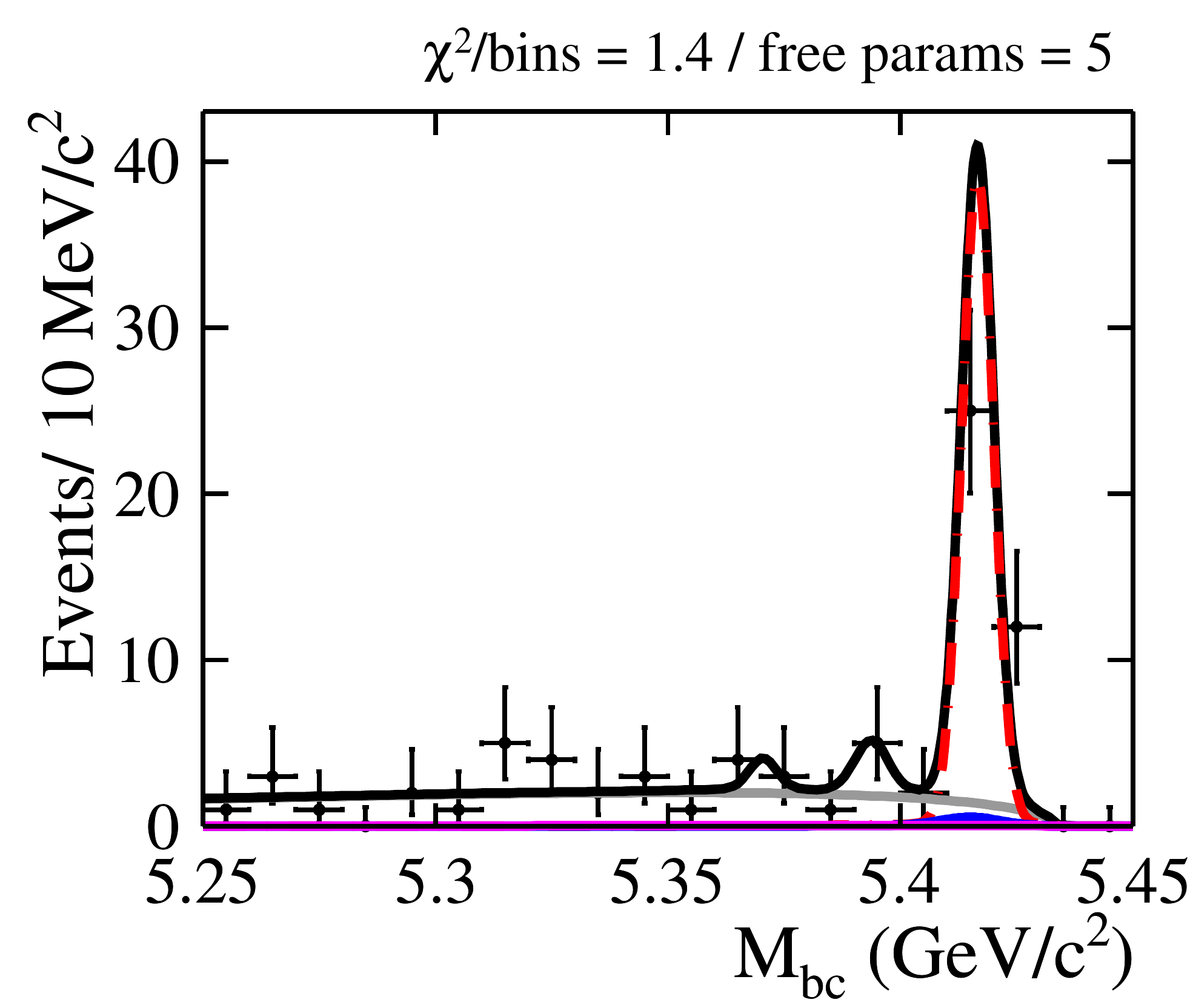}
\includegraphics[width=55mm]{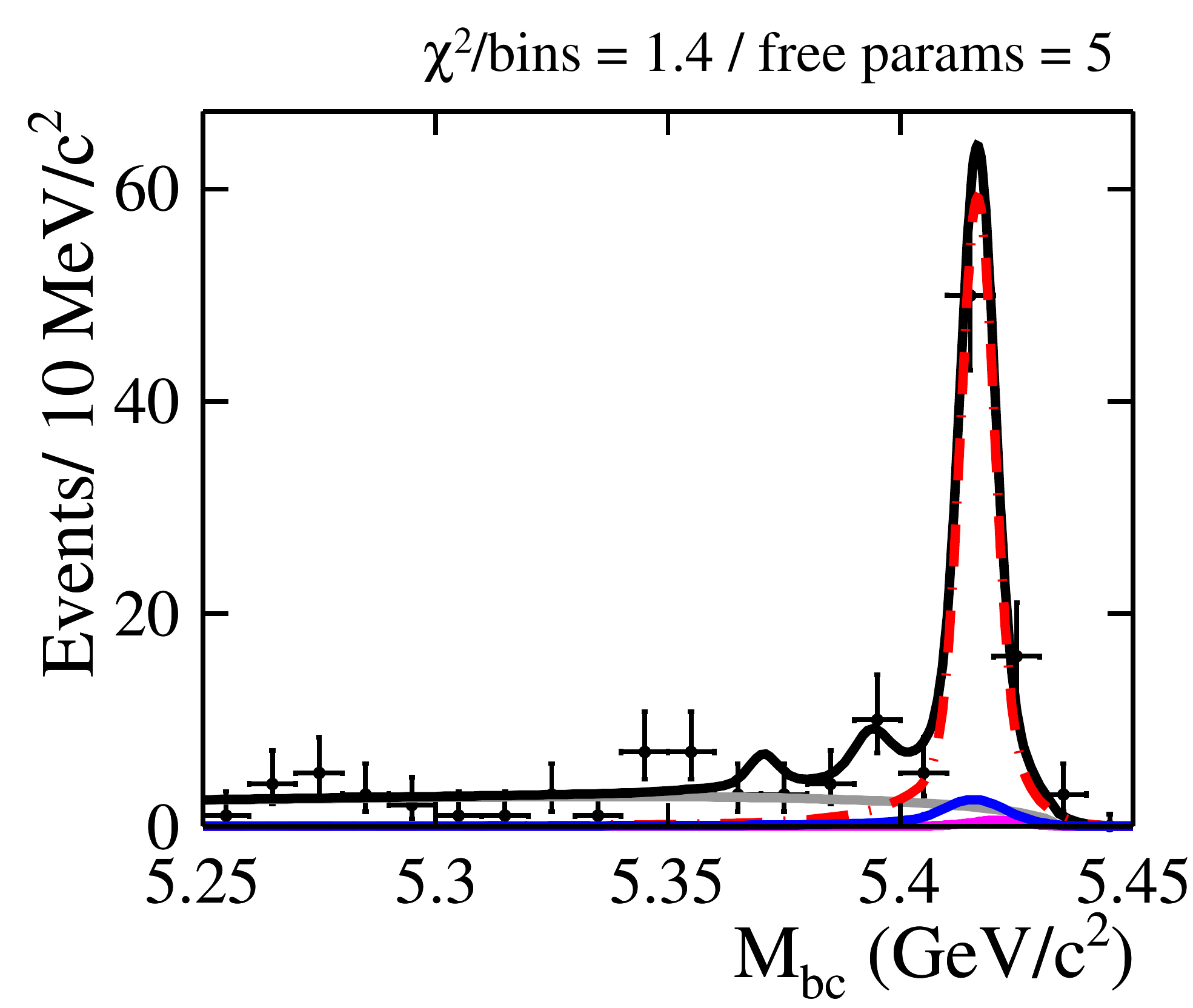}
\includegraphics[width=55mm]{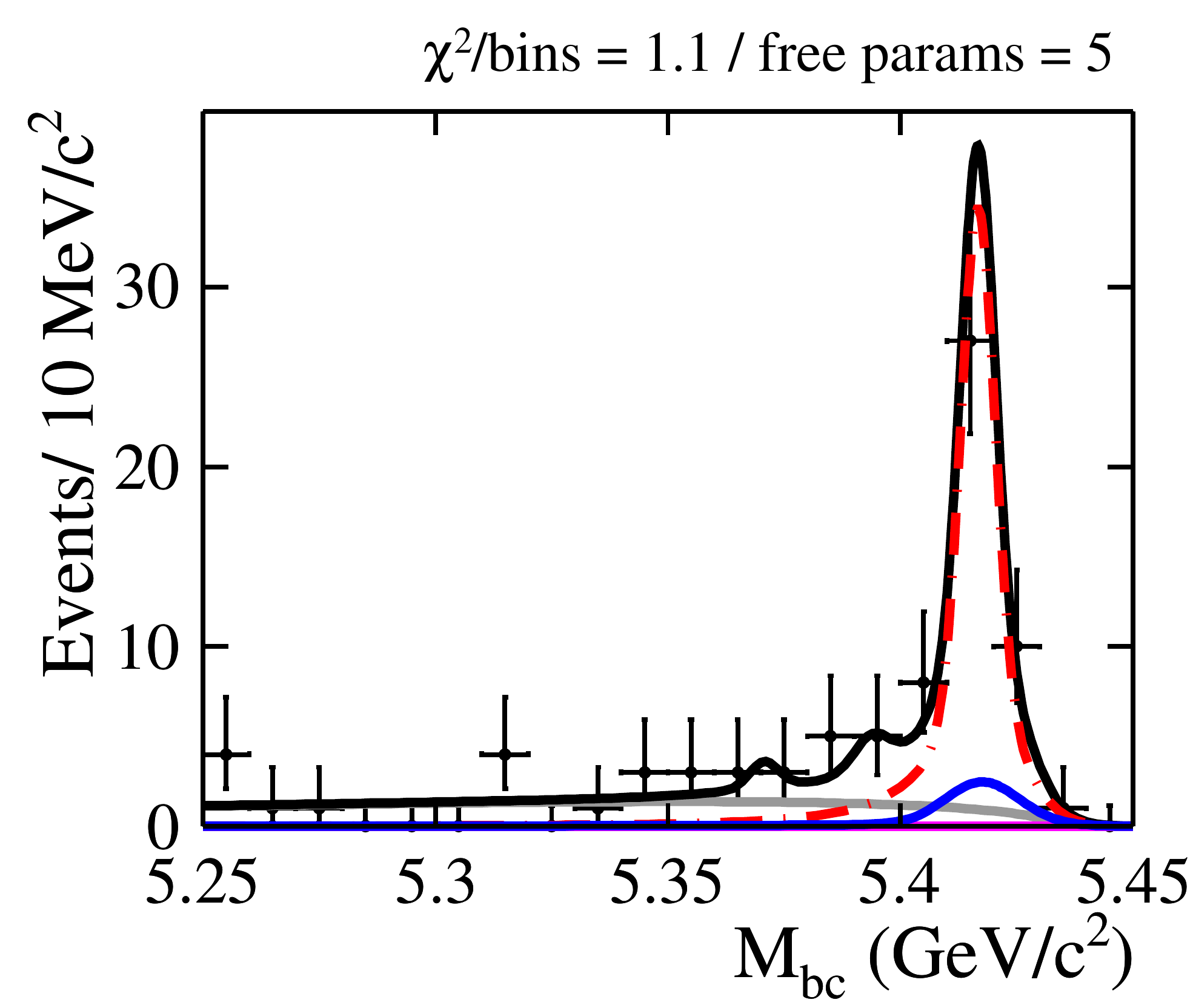}

\caption{$\mbc$ and $\de$ projections of the fit result. The columns correspond to $B^0_s\ra D^+_s D^-_s$ (right), $B^0_s\ra D^{*\pm}_s D^{\mp}_s$ (middle), and $B^0_s\ra D^{*+}_s D^{*-}_s$ (left). The red dashed curves show RC+WC signal, the blue-purple solid curves show CF, the grey solid curve shows background, and the black solid curves show the total.} \label{fig:fit_results}
\end{figure}

We measure the signal yields for \bsdsds\ decays using $7.1\pm1.3$ million $B_s^{(*)}\bar{B}_s^{(*)}$ pairs with a $\bsst\bsbarst$ fraction $f^{}_{\bsst\bsbarst}=(87.0\pm1.7)\%$~\cite{remi} . 
The fit results are listed in Table~\ref{tab:fit_results}, and projections of the fit are shown in Fig.~\ref{fig:fit_results}. The branching fraction for channel $i$ is calculated as ${\cal B}^{}_i = Y^{}_i/(\varepsilon^i_{MC}\cdot N^{}_{B_s^{(*)}\bar{B}_s^{(*)}} \cdot f^{}_{\bsst\bsbarst} \cdot 2)$, where $Y^{}_i$ is the fitted CR yield, and $\varepsilon^i_{MC}$ is the MC efficiency with intermediate branching fractions~\cite{pdg} included. The efficiencies $\varepsilon^i_{MC}$ include small  correction factors to account for differences between MC and data for kaon identification.

The systematic errors are listed in Table~\ref{tab:syst_errors}. The error due to PDF shapes is evaluated by varying shape parameters by $\pm 1\sigma$. The systematic error for the fixed WC and CF-down  fractions is evaluated by repeating the fit with each fixed fraction varied by $\pm 20$\%.
The uncertainties due to $K^\pm$ identification and tracking are $\sim$\,1\%(momentum-dependent) and 0.35\% per track respectively. 
As the longitudinal polarization fraction ($\fl$) of $B^0_s\ra D^{*+}_sD^{*-}_s$ is not measured yet, we assume $\fl$ to be the world average (WA) value for the analogous spectator decay $B^0_d\ra D^{*+}_s D^{*-}$: $0.52\pm\,0.05$~\cite{pdg}. 
The related systematic error  is taken as the change in ${\cal B}$ when $f^{}_L$ is  varied by twice the error on the WA value.
Significant uncertainties arise from $D^+_s$ branching fractions and the fraction of $\Upsilon(5S)$ decays producing $\bs$ mesons, which are external factors that are expected to be measured more precisely in the future. 
The statistical significance given in Table~\ref{tab:fit_results} is calculated as $\sqrt{-2\ln(\mathcal{L}_0 / \mathcal{L}_{\mathrm{max}})}$, where $\mathcal{L}_0$ and $\mathcal{L}_{\mathrm{max}}$ are the values of the likelihood function when the signal yield $Y^{}_i$ is fixed to zero and when it is set to the fitted value, respectively. We include systematic uncertainty in the significance by smearing the likelihood function by a Gaussian having a width equal to the total systematic error obtained for the signal yield.

\begin{table*}[ptbh]
\caption{Signal yield ($Y$), efficiency including intermediate branching fractions ($\varepsilon$), branching fraction (${\cal B}$), and signal significance ($S$) including systematic uncertainty. The first errors listed are statistical, and the second are systematic. }
\begin{center}
\renewcommand{\arraystretch}{1.4}
\begin{tabular*}{0.9\textwidth}
{@{\extracolsep{\fill}} l | c c c c }
\hline\hline
Mode & $Y$ & $\varepsilon$ & ${\cal B}$ & $S$ \\
 & (events) & ($\times 10^{-4}$) & (\%) &   \\
\hline
$D^+_s D^-_s$ & 
 $33.1_{-5.4}^{+6.0}$ & 4.72 &  $0.58^{+0.11}_{-0.09}$ $\pm0.13$ &11.5\\  [1.0ex] 
$D^{*\pm}_s D^{\mp}_s$ & 
$44.5_{-5.5}^{+5.8}$ & 2.08 &  $1.8\pm0.2$ $\pm0.4$ &10.1\\  [1.0ex] 
$D^{*}_s D^{*}_s$ & 
$24.4_{-3.8}^{+4.1}$ & 1.01 &  $2.0\pm0.3$ $\pm0.5$ &7.8\\  [1.0ex] 

\hline
Sum & 
$102.0_{-8.6}^{+9.3}$ &  &  $4.3\pm0.4$ $\pm1.0$ & \\  [0.5ex] 

\hline\hline
\end{tabular*} 
\end{center}
\label{tab:fit_results}
\end{table*}
\begin{table*}[ptbh]

\caption{\label{tab:syst_errors}Systematic errors (\%). Those listed in the top section affect the signal yield and thus the signal significance.}
\centering
\renewcommand{\arraystretch}{1.2}
\begin{tabular*}{0.9\textwidth}{@{\extracolsep{\fill}} l  | c c c c  c c @{\extracolsep{\fill}} }
\hline
\hline
Source & 
\multicolumn{2}{c}{$D^+_sD^-_s$} & 
\multicolumn{2}{c}{$D^*_s D^{}_s$} & 
\multicolumn{2}{c}{$D^{*+}_s D^{*-}_s$} \\
\hline
 & $+\sigma$ & $-\sigma$ & $+\sigma$ & $-\sigma$ & $+\sigma$ & $-\sigma$ \\
\hline
Signal PDF Shape   & 2.7 &2.2   & 2.2& 2.4& 5.1& 3.8 \\
Background PDF  Shape & 1.5 &1.2   & 1.3& 1.4& 2.9& 2.2 \\
WC + CF fraction     & 0.7 & 0.6& 4.6 & 4.5 & 6.2 & 6.2 \\
$\mathcal{R}$ requirement ($q\bar{q}$ suppression)  & 	3.1 & 0.0	& 0.0 & 2.7 & 0.0 & 2.1   \\
Best candidate selection & 5.5 & 0.0	& 1.5 & 0.0 & 1.5 &  0.0   \\
$K^{\pm}$ Identification      & 7.0 & 7.0	& 7.0 & 7.0 & 7.0 & 7.0  \\
$K_{S}$  Reconstruction      & 1.1   & 1.1 & 1.1 & 1.1 & 1.1& 1.1  \\
 $\pi^{0}$ Reconstruction & 1.1 & 1.1 & 1.1 & 1.1 & 1.1 & 1.1 \\
$\gamma$& - &-	& 3.8 & 3.8 & 7.6 & 7.6 \\
Tracking    &  2.2 & 2.2 & 2.2 & 2.2 & 2.2	& 2.2  \\
Polarization   &  0.1 & 0.1 & 0.8 & 0.7 & 0.5	& 1.0  \\
\hline
MC statistics for $\varepsilon$ & 0.2 &  0.2	& 0.4  & 0.4 &  0.5 &  0.5 \\
$D_{s}^{(*)}$ Branching Fractions  & 8.6& 8.6& 8.6&8.6& 	8.7& 8.7 \\
$N_{B_s^{(*)}B_s^{(*)}}$            & \multicolumn{6}{c}{18.3} \\
$f^{}_{B^*_s\overline{B}^*_s}$ & \multicolumn{6}{c}{2.0} \\
\hline
Total  & 22.7 & 21.8  &  22.6 & 22.8 & 24.6 & 24.3 \\
\hline
\hline
\end{tabular*}

\end{table*}


In the heavy quark limit with $(m^{}_b-2m^{}_c)\ra 0$ and  $N^{}_c\ra\infty$, the dominant contribution to the decay width comes from \bsdsds\ decays~\cite{Shifman,Aleksan}.
Assuming negligible \cp\ violation, the branching fraction is related to $\dgs$ as $\dgs/\gs = 2{\cal B}/(1-{\cal B})$.
Inserting the total ${\cal B}$ from Table~\ref{tab:fit_results} gives
\begin{eqnarray}
\frac{\dgs}{\gs} & = & 
0.090\pm0.009 \,\pm0.022\,,
\label{eqn:dg_result}
\end{eqnarray}
where the first error is statistical and the second is systematic. 
This result is in good agreement with the current WA~\cite{pdg} and is consistent with theory~\cite{Nierste}. There is a theoretical uncertainty arising mainly from the \cp-odd component in  $B^0\ra  D^{*+}_sD^{*-}_s$ and the unknown contribution of 3-body final states.

\begin{figure}[ptbh]
\centering
\includegraphics[width=100mm]{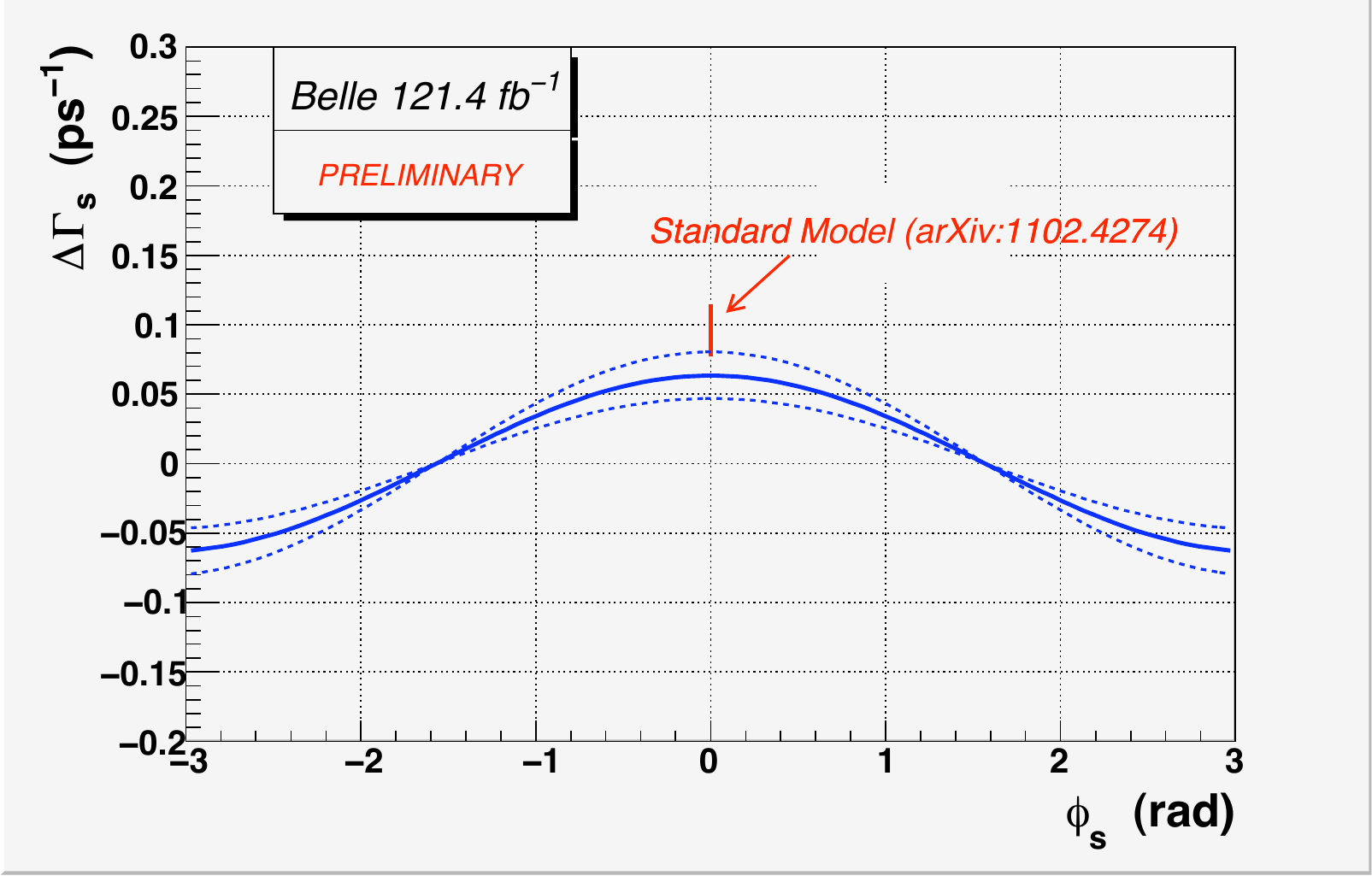}
\caption{ The width difference $\dgs$ as a function of $\phi_s$. One-sigma band and SM value are shown for comparison.  }
 \label{fig:dgs}
\end{figure}

If a \cp -violating phase $\phi_s$ is allowed, the above relation becomes 
\begin{eqnarray}
4{\cal B}(B^0_s\ra D^{(*)}_sD^{(*)}_s) = \left(\frac{\dgcp}{\cos\phi_s}\right)\left[\frac{1+\cos\phi_s}{1+\dgcp}+\frac{1-\cos\phi_s}{1-\dgcp} \right]\,,
\label{eqn:dg_w_cpv}
\end{eqnarray}

where $\phi_s=Arg(M_{12}/\Gamma_{12})$~\cite{Dunietz}. Fig.~\ref{fig:dgs} plots $\dgs$ as a function of $\phi_s$ for our measurement.

In summary, we have measured the branching fractions for \bsdsds\ using $e^+e^-$ data taken at the $\Upsilon(5S)$ resonance. Our results constitute the first observation of $B^0\ra D^{*\pm}_s D^{*\mp}_s$  ($8\sigma$ significance).  Using the total measured branching fraction ${\cal B}(B^0_s\ra D^{(*)+}_s D^{(*)-}_s)=(4.3\,\pm 0.4\,\pm 1.0)\%$ and assuming no \cp\ violation, we determine the relative $\bs\bsbar$ decay width difference to be ~$\dgs/\gs=0.090\pm0.009 \,\pm0.022$.


\bigskip 

\begin{thebibliography}{99} 

\bibitem{bs_cleo} 
G.\,S.\,Huang {\it et al.} (CLEO Collab.), Phys.\,Rev.~D {\bf 75}, 012002 (2007). 

M.~Artuso {\it et al.} (CLEO Collab.), Phys.\,Rev.\,Lett.~{\bf 95}, 261801 (2005). 

\bibitem{drutskoy07} 
A.~Drutskoy {\it et al.} (Belle Collab.), Phys.\,Rev.\,Lett.~{\bf 98}, 052001 (2007). 


\bibitem{belle_detector} A.~Abashian {\it et al.\/} (Belle Collab.), 
Nucl. Instr. Meth. Phys. Res.~A {\bf 479}, 117 (2002).

\bibitem{kekb} S.~Kurokawa and E.~Kikutani, Nucl. Instrum. and 
Methods Phys. Res.~A {\bf 499}, 1 (2003), and other papers included in this volume.


\bibitem{charge-conjugates} 
Charge-conjugate modes are implicitly included.

\bibitem{Aleksan} R. Aleksan {\it et al.\/}, Phys. Lett.~B {\bf 316}, 567 (1993).

\bibitem{esen10} 
S.~Esen {\it et al.} (Belle Collab.), Phys.\,Rev.\,Lett.~{\bf 105}, 201802 (2010). 

\bibitem{pdg} K. Nakamura {\it et al.} (Particle Data Group),
 Jour.\, of Phys.~G {\bf 37}, 075021 (2010) and 2011 partial update for the 2012 edition.

\bibitem{goodKS} 
Y.~Nakahama {\it et al.\/} (Belle Collab.), 
Phys. Rev. Lett.~{\bf 100}, 121601 (2008).

\bibitem{KSFW} G. C. Fox and S. Wolfram, 
Phys. Rev. Lett.~{\bf 41}, 1581 (1978). 
The modified moments used in this paper are described in 
S.\,H.\,Lee {\it et al.} (Belle Collab.), 
Phys. Rev. Lett.~{\bf 91}, 261801 (2003).

\bibitem{ARGUS}
H.~Albrecht {\it et al.\/} (ARGUS Collab.), 
Phys. Lett.~B {\bf 241}, 278 (1990). 

\bibitem{fit_errors} Thus the fit errors for yields are less than 
the square root of the yields due to the CF information. 

\bibitem{remi} R.~Louvot {\it et al.} (Belle Collab.), 
Phys.\,Rev.\,Lett.~{\bf 102}, 021801 (2009). 

\bibitem{Shifman} M.\,A.\,Shifman and M.\,B.\,Voloshin,
Sov. J. Nucl. Phys.~{\bf 47}, 511 (1988).


\bibitem{Nierste} A.~Lenz and U.~Nierste, 
arXiv:1102.4274;


\bibitem{Dunietz} I.\ Dunietz, R.\ Fleischer, and U.\ Nierste, 
  Phys.\ Rev.~D {\bf 63}, 114015 (2001);
  I.\ Dunietz, Phys.\ Rev.~D {\bf 52}, 3048 (1995).


\end{thebibliography}

\end{document}